\documentstyle[aps,psfig]{revtex}
\begin{document}
\draft \twocolumn

\title{Bichromatic electromagnetically induced transparency in cold rubidium atoms}

\author{J. Wang$^{1,2\ast}$, Yifu Zhu$^{3,2}$, K. J. Jiang$^{1,2}$,and M. S. Zhan$^{1,2}$}

\address{$^{1}$State Key Laboratory of Magnetic Resonance and Atomic
and Molecular Physics, Wuhan Institute of Physics and Mathematics,
Chinese Academy of Sciences, Wuhan 430071, China}
\address{$^{2}$Center for Cold Atom Physics, Chinese Academy of Sciences,
Wuhan 430071, China }
\address{$^{3}$Department of Physics,
Florida International University, Miami, Florida 33199, USA}

\date{\today}

\maketitle

\begin{abstract}
{\it In a three-level atomic system coupled by two equal-amplitude
laser fields with a frequency separation 2$\delta$, a weak probe
field exhibits a multiple-peaked absorption spectrum with a
constant peak separation $\delta$. The corresponding probe
dispersion exhibits steep normal dispersion near the minimum
absorption between the multiple absorption peaks, which leads to
simultaneous slow group velocities for probe photons at multiple
frequencies separated by $\delta$. We report an experimental study
in such a bichromatically coupled three-level $\Lambda$ system in
cold $^{87}$Rb atoms. The multiple-peaked probe absorption spectra
under various experimental conditions have been observed and
compared with the theoretical calculations.}
\end{abstract}



A resonant laser beam can pass through an opaque atomic medium
without attenuation due to the quantum interference effect between
the dressed states created by a coupling laser field. This
phenomenon is known as electromagnetically induced transparency
(EIT) \cite{1,2,3}. In recent years, many studies on EIT and
related phenomena have been carried out, which reveal the
importance of EIT in understanding the fundamental physics
involving interactions between light field and resonant medium
\cite{4,5,6,7,8}. It has been shown that EIT may have applications
in a variety of research topics such as quantum optics with slow
photons \cite{9,10,11,12,13}, quantum information processing
\cite{14}, atomic frequency standard \cite{15,16,17,18}, and
quantum nonlinear optics \cite{19,20}.

Recently, Lukin et al proposed a mechanism to entangle two photons
in an EIT medium based on obtaining slow photons at different
frequencies \cite{21}. Since the EIT created by a monochromatic
field only provides the steep dispersion near the resonant
frequency, sophisticated schemes are proposed to obtain slow
photons at different frequencies \cite{21,22}. Here, we show that
EIT in a $\Lambda$ type level configuration created by a
bichromatic laser field may be used to slow down photons at
different frequencies. The three-level $\Lambda$ system coupled by
a bichromatic field and a probe field is depicted in Fig. 1(a).
The dressed states created by the bichromatic field consist of an
infinite ladder with an equal-spacing separation $\delta$ between
the neighboring levels when the average frequency of the
bichromatic field with equal amplitudes of the two frequency
components matches the atomic transition frequency. The dressed
states are the superposition of the atomic states
$\vert$%
2%
$>$%
,
$\vert$%
3%
$>$%
, and the photon number states with the amplitude determined by
the Rabi frequency ($\Omega_{c}=\Omega_{c1}=\Omega_{c2}$) and the
frequency separation $2\delta$. Such dressed states and the
fluorescence spectrum of the two-level atoms coupled by a
bichromatic field have been extensively studied before
\cite{23,24,25,26}. The dressed state picture of the bichromatic
driven three-level system is depicted in Fig. 1(b). It is expected
that the probe absorption spectrum will exhibit multiple peaks
corresponding to the dressed transitions
$|1>\rightarrow|\mathit{m}>$ and transparent windows with minimum
absorption located near the middle separation of the dressed
states. In particular, when $\Omega<\delta$, the transition
amplitudes will be dominant only for a few dressed state around
$\mathit{m}=0$.

\begin{figure}[tb]
\psfig{figure=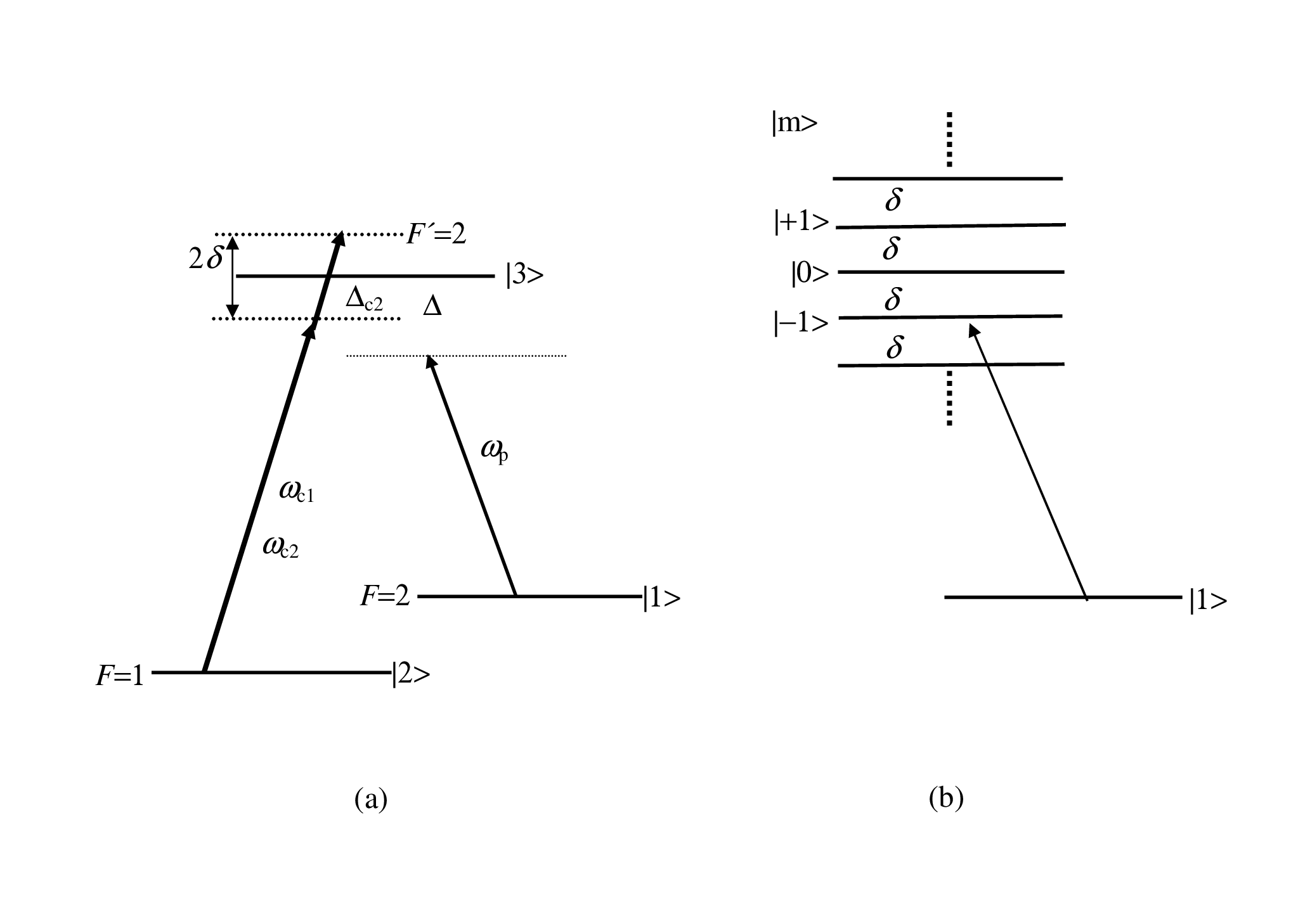,width=\columnwidth,angle=270} \caption{ (a)
Three-level $\Lambda$ type system coupled by a bichromatic field
and a weak probe field. (b) Dressed state picture of the coupled
three-level system. $\omega_{c1}$ and $\omega_{c2}$ are the
frequencies of the bichromatic field. The frequency detuning
$\Delta_{c2}=\omega_{0}-\omega_{c2}$ ($\omega_{0}$ is the atomic
transition frequency). $\omega_{p}$ is the probe laser frequency.
$2\delta=\omega_{c1}-\omega_{c2}$ is the frequency difference of
the bichromatic field.}
\end{figure}

We have numerically solved the density matrix equations of the
bichromatic coupled three-level $\Lambda$ system with a continued
fraction \cite{27} method and the results are plotted in Fig. 2.
Fig. 2(a) presents the calculated probe absorption and dispersion
for a moderate coupling Rabi frequency
($\Omega_{c}=\Omega_{c1}=\Omega_{c2}=0.4\Gamma$) and the frequency
separation of the bichromatic components $2\delta=1.4\Gamma.$The
absorption spectrum exhibits three peaks, corresponding to the
dressed state transition
$\vert$%
1%
$>$
to
$\vert$%
$\mathit{m}=0$%
$>$
and $|\mathit{m}=\pm1>$. The probe dispersion exhibits a normal
steep slope at the absorption minimum near $\pm\delta/2$. This
implies that the bichromatic EIT medium supports slow photons at
frequencies $\omega_{p}\pm\delta/2$ simultaneously. In Fig. 2(b),
we plot the calculated probe absorption and dispersion for a
strong bichromatic field. The spectrum exhibits more peaks
indicating that the transition amplitudes spread to more dressed
states. The probe dispersion again shows the normal slope at the
absorption minimum, which can be used to obtain slow group
velocities simultaneously for photons with different frequencies.
Our calculations show that for the optimal performance of two slow
photons at frequencies $\omega_{p}\pm\delta/2$, it is desirable to
have a moderate Rabi frequency $\Omega<\Gamma$ and $2\delta\sim
\Gamma $( in order to optimize the slope of the normal dispersion
and the = dispersion amplitude).Then the frequency bandwidth of
the light pulses $\Delta\nu$ should satisfy $\Delta\nu<\delta.$For
comparison, Fig. 2(c) shows the monochromatic EIT spectrum and the
corresponding dispersion curve under similar conditions.

\begin{figure}[tb]
\psfig{figure=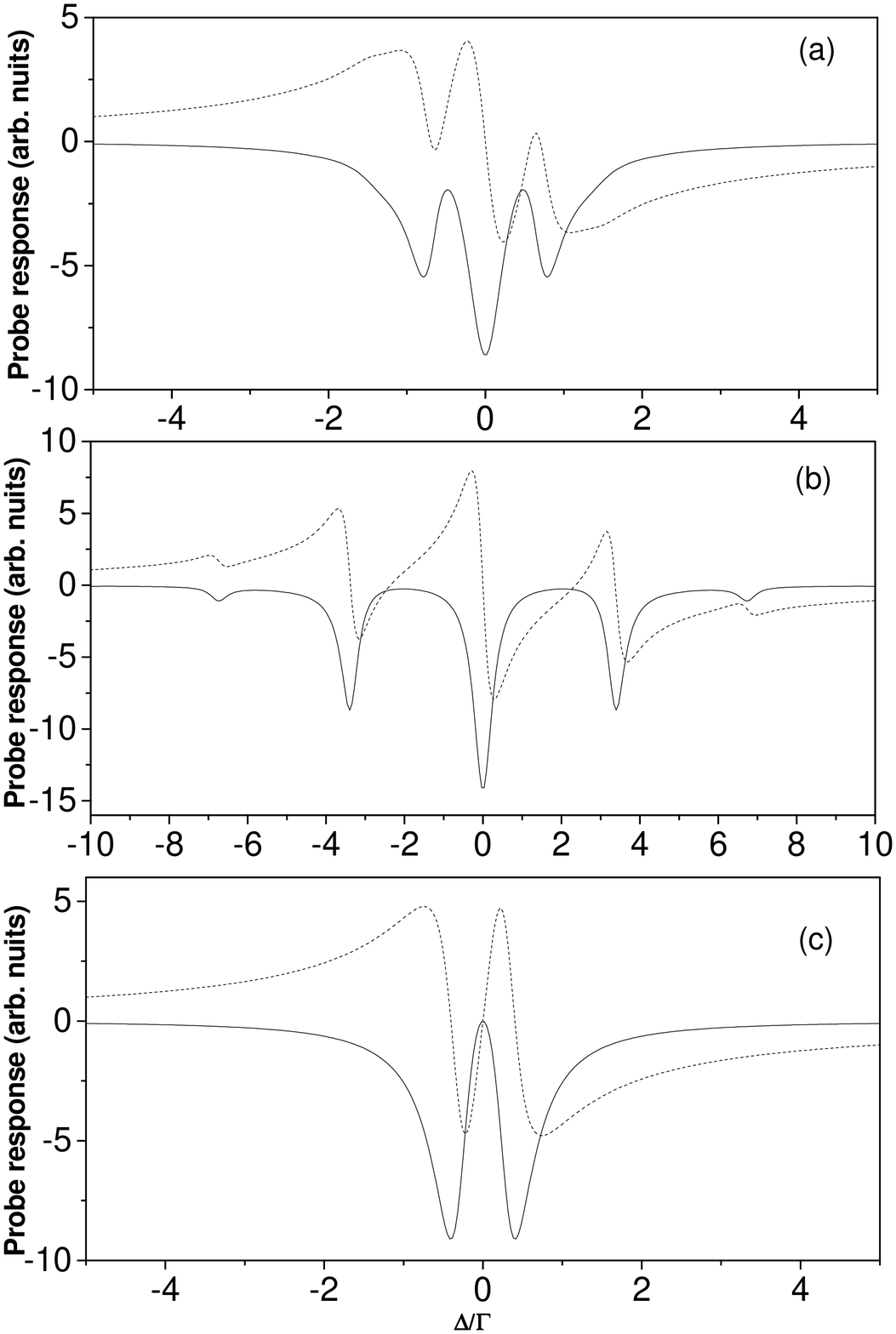,width=0.7\columnwidth,angle=0} \caption{
Calculated probe response in the three-level $\Lambda$ system
exhibiting bichromatic EIT. Solid line is the probe absorption and
dash line is the probe dispersion. $\Gamma$ is the decay rate of
the excited state. (a) $\Omega_{c1}=\Omega_{c2}=0.4\Gamma$ and
$2\delta=1.4\Gamma$. (b) $\Omega_{c1}=\Omega_{c2}=2\Gamma$ and
$2\delta=6.7\Gamma$. (c) The probe response in the identical
three-level $\Lambda$ system with a monochromatic coupling field
($\Omega_{c}=0.4\Gamma$).}
\end{figure}

We have performed an experimental study of bichromatic
electromagnetically induced transparency in a three-level
$\Lambda$ system of cold $^{87}$Rb atoms confined in a MOT
\cite{28}. The frequency separation of the bichromatic coupling
field is 2$\delta$ (taken as 40 MHz or 80 MHz). The observed probe
absorption spectrum is qualitatively different from the spectrum
of the usual three-level $\Lambda$ EIT with a monochromatic
coupling field. There are several absorption peaks and
transparency dips in bichromatic EIT profile separated in
frequency by $\delta$. We studied the dependence of probe spectrum
on the coupling field intensity and the frequency separation and
found that the experimental results agree with the theoretical
calculations.

\begin{figure}[tb]
\psfig{figure=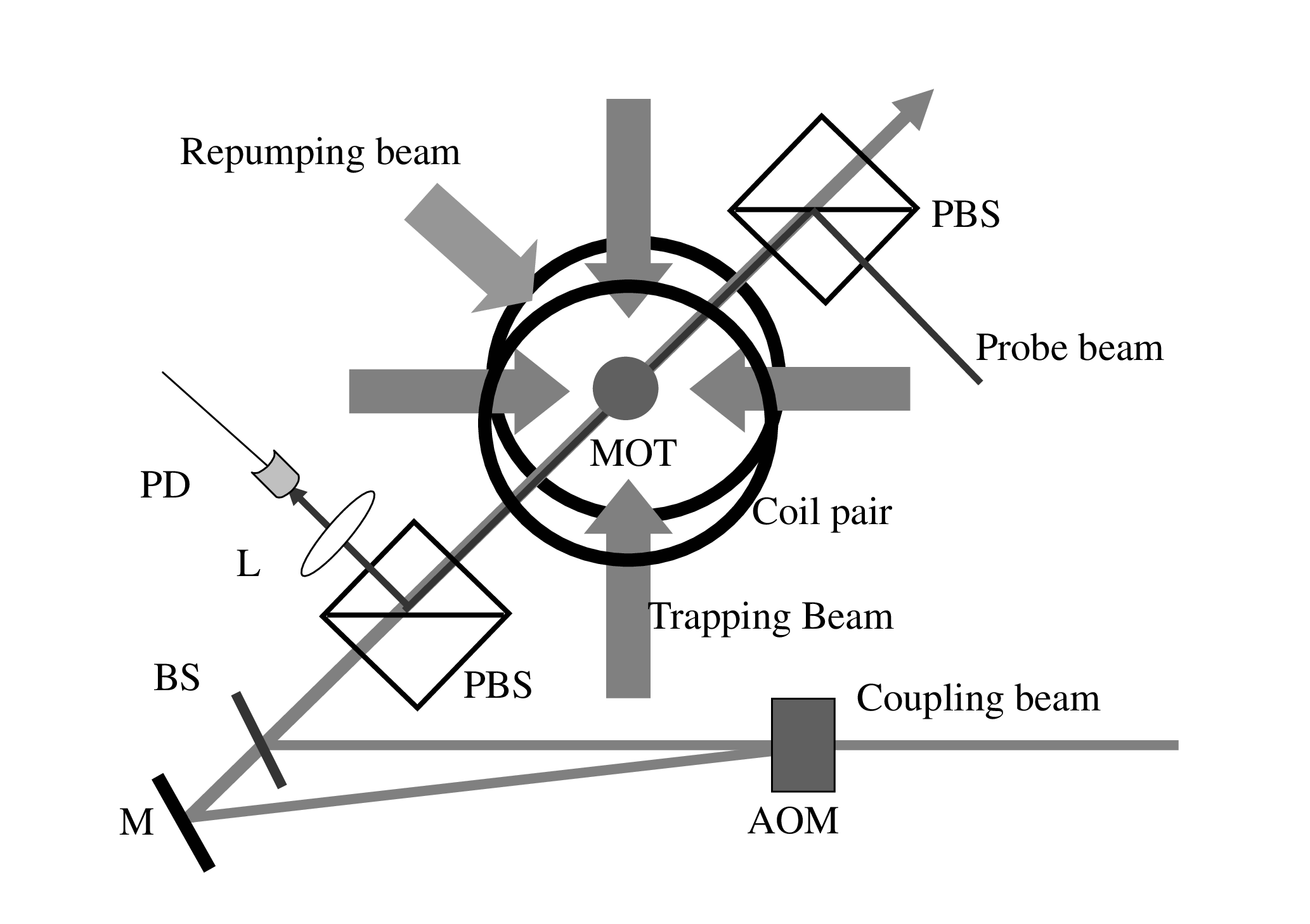,width=\columnwidth,angle=270} \caption{
Schematic diagram of the experimental setup. PBS: polarizing beam
splitter; PD: photodiode; L: lens; BS: beam splitter; M: mirror;
AOM: acousto-optic modulator.}
\end{figure}

The three-level $\Lambda$ type EIT system of $^{87}$Rb atoms is
shown in Fig. 1(a). A bichromatic coupling laser drives the
\textit{D}$_{2}$ transition $5\mathit{S}_{1/2}$,
$\mathit{F}=1\rightarrow5\mathit{P}_{3/2},\mathit{F}%
\acute{}%
=2$ and a weak probe laser with frequency $\omega_{p}$ drives the
transition
$5\mathit{S}_{1/2},\mathit{F}=2\rightarrow5\mathit{P}_{3/2}$, =
$\mathit{F}%
\acute{}%
=2$. The bichromatic coupling field is produced by combining the
diffracted zeroth-order and the first-order beams from a single
laser beam passing through an acousto-optic modulator (AOM). The
two frequencies of the bichromatic field are
$\omega_{c1}=\omega_{c}+\delta$ and $\omega_{c2}%
=\omega_{c}-\delta$ with the respective Rabi frequencies $\Omega_{c1}%
=\Omega_{c2}=\Omega_{c}$. A simplified diagram of the experimental
apparatus is depicted in Fig. 3. A homemade octagon quartz cell is
used as the MOT cell. The inner pressure of the cell is
$2\times10^{-7}$ Pa . The cooling and trapping beam (at 780 nm) is
provided by a tapered amplifier diode laser (TOPTICA TA100) and
the laser frequency is stabilized by the saturated absorption
spectroscopic method. The repumping laser (at 780nm) is provided
by an extended-cavity diode laser (TOPTICA DL100). The Coupling
laser for the bichromatic EIT study is provided by a Ti: sapphire
laser (Coherent MBR 110) with a beam diameter $\sim$3 mm. An AOM
is used to produce the zeroth-order, and the frequency shifted
first-order beams and followed by a mirror and a beam splitter to
combine the two beams. Another extended-cavity diode laser
(TOPTICA DL100) provides the weak probe beam with a diameter
$\sim$1 mm. The coupling and probe laser beams are linearly
polarized perpendicular to each other. They pass through the MOT
in opposite directions and overlap in the path via the
polarization beam splitters (PBS). A photodiode (PD) and a digital
oscilloscope (Tektronix TDS 220) are used to detect and record the
probe attenuation signal.

In the experiment, the trapping laser frequency is red-detuned and
then locked \cite{29} by an amount $\sim-2\Gamma$ relative to the
resonant frequency of the $5\mathit{S}_{1/2}$,
$\mathit{F}=2\rightarrow5\mathit{P}%
_{3/2},\mathit{\mathit{F}%
\acute{}%
=}3$ transition and the repumping laser frequency is locked to the
$5\mathit{S}_{1/2},\mathit{F}=1\rightarrow5\mathit{P}_{3/2}$,
$\mathit{\mathit{F}%
\acute{}%
=}2$ transition. A near-spherical $^{87}$Rb atom cloud with a
diameter $\sim$3 mm is formed that contains about $5\times10^{7}$
atoms with the temperature of the atom cloud $\sim100$ $\mu$K. We
adjusted the steering mirror of probe beam to optimize the spatial
overlap of the probe beam and the cold atom cloud by monitoring
the probe absorption amplitude. The maximum absorption of $\sim
30\% $ is obtained when we tune the probe
frequency to the peak of the $\mathit{D}_{2}$ transition $5\mathit{S}%
_{1/2},\mathit{F}=2\rightarrow5\mathit{P}_{3/2},\mathit{F}%
\acute{}%
=2$. Then, the bichromatic coupling laser is applied and
overlapped with the probe beam. The intensities of bichromatic
beams are balanced via neutral density filters placed before the
combing beam splitter. During the experiment, the bichromatic
coupling laser is tuned to the $\mathit{D}_{2}$
line transition $5\mathit{S}_{1/2},\mathit{F}=1\rightarrow5\mathit{P}%
_{3/2},\mathit{F}%
\acute{}%
=2$, the probe frequency is scanned across the $5\mathit{S}_{1/2}%
,\mathit{F}=2\rightarrow5\mathit{P}_{3/2}$, $\mathit{F}%
\acute{}%
=3,$transition, and the transmitted probe light is detected and
recorded. Because the two coupling laser beams have a fixed
frequency difference, we define the coupling detuning as
$\Delta_{c2}=\omega_{0}-\omega_{c2}$, where $\omega_{0}$ is the
resonant frequency of the $^{87}$Rb \textit{D}$_{2}$ transition
$5\mathit{S}_{1/2},\mathit{F}=1\rightarrow5\mathit{P}_{3/2}$,
$\mathit{F}%
\acute{}%
=2,$ and $\omega_{c2}$ is the frequency of the red-detuned
component of the bichromatic coupling beams.

\begin{figure}[tb]
\psfig{figure=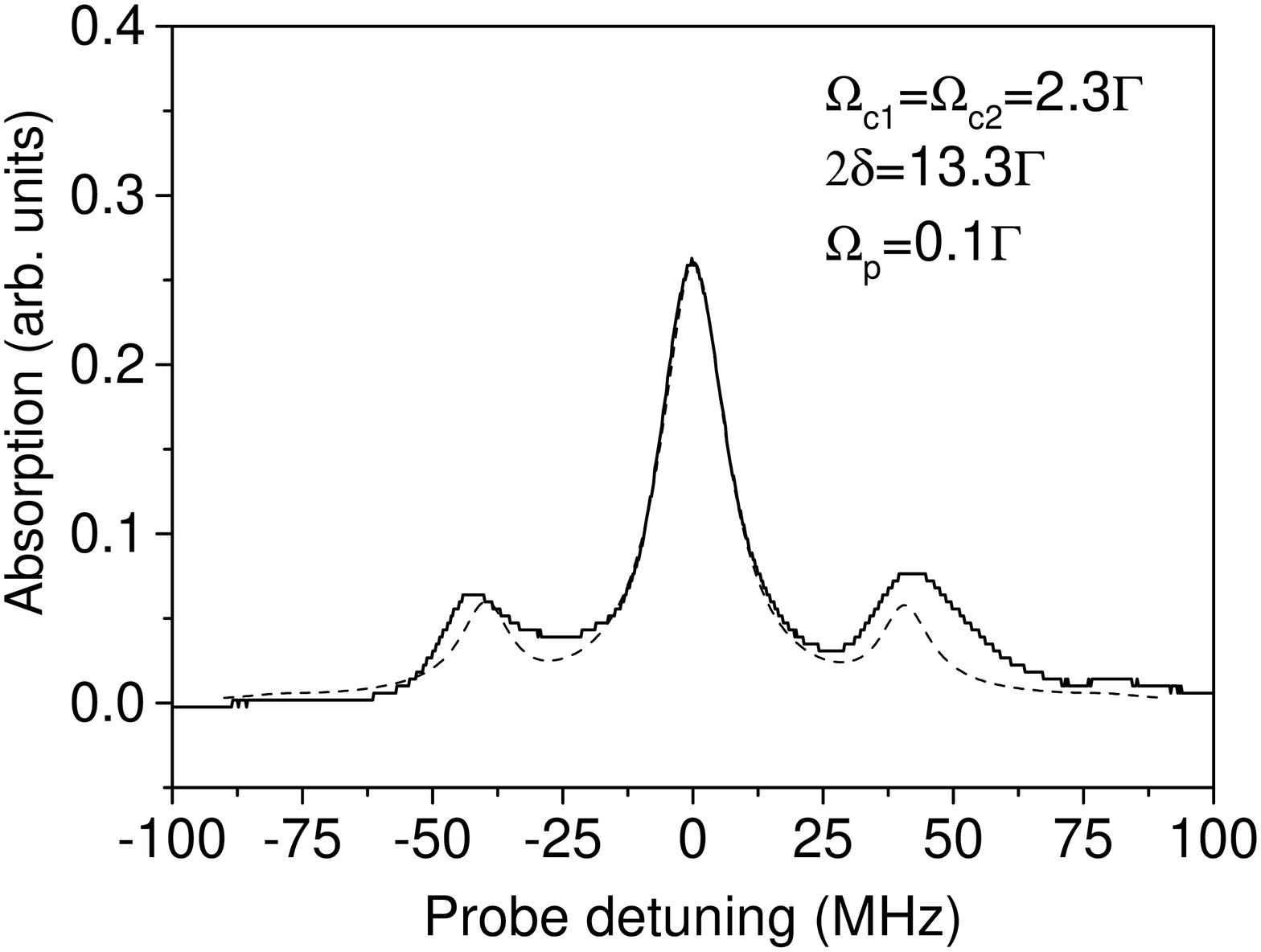,width=0.9\columnwidth,angle=270}
\caption{Measured probe absorption versus the probe detuning in
the bichromatic coupled three-level $^{87}$Rb atoms. The solid
line is the experimental data and the dash line is the calculated
fit. The frequency difference $2\delta=80$ MHz; the coupling Rabi
frequencies $\omega_{c1}=\omega_{c2}=2.3\Gamma$ ($\Gamma=6$ MHz);
and the Rabi frequency of probe field $\Omega_{c}=0.1\Gamma $.}
\end{figure}

Fig. 4 shows the probe absorption spectrum recorded with the
frequency separation $2\delta=80$ MHz and the coupling Rabi
frequency $\Omega _{c1}=\Omega_{c2}=14$ MHz (the solid line). The
average frequency of the bichromatic field is resonant with the
atomic transition ($\Delta_{c2}=\delta $). The symmetrical
absorption spectrum exhibits three absorption peaks with the
neighboring peak separation equal to $\delta$, the half frequency
difference of the bichromatic coupling field. The dashed line in
Fig. 4 represents the theoretical fit. The linewidth of the
absorption peak is broadened by the Zeeman shifts (the magnetic
field is on during the experimental process), the induced decay of
the ground state coherence due to the presence of the MOT lasers,
and the finite laser linewidth. The calculated results are
obtained with a suitable line average to account for the Zeeman
spectral broadening.

To produce a bichromatic field with $\delta=20$ MHz, we use two
AOMs (not shown in Fig. 4) with 80 MHz and 120 MHz carrier
frequencies and combine the 1st-order diffracted beams of the two
AOMs as the bichromatic coupling field. The measured probe
absorption spectra are presented in Fig. 5 and 6. The solid lines
are the experimental data and dash lines are the theoretical fits.
There are more peaks and EIT dips, differing from the probe
absorption profile obtained with $\delta=40$ MHz. The Bichromatic
EIT profile also changes with the coupling beam intensity: more
peaks appear with stronger coupling field. Fig. 5(a), 5(b), 5(c),
and 5(d) are measured with a resonant bichromatic field
($\Delta_{c2}=\delta$), and the coupling Rabi frequencies
$\Omega_{c1}%
=\Omega_{c2}=14$ MHz, 12 MHz, 10 MHz, and 9 MHz respectively.

\begin{figure}[tb]
\psfig{figure=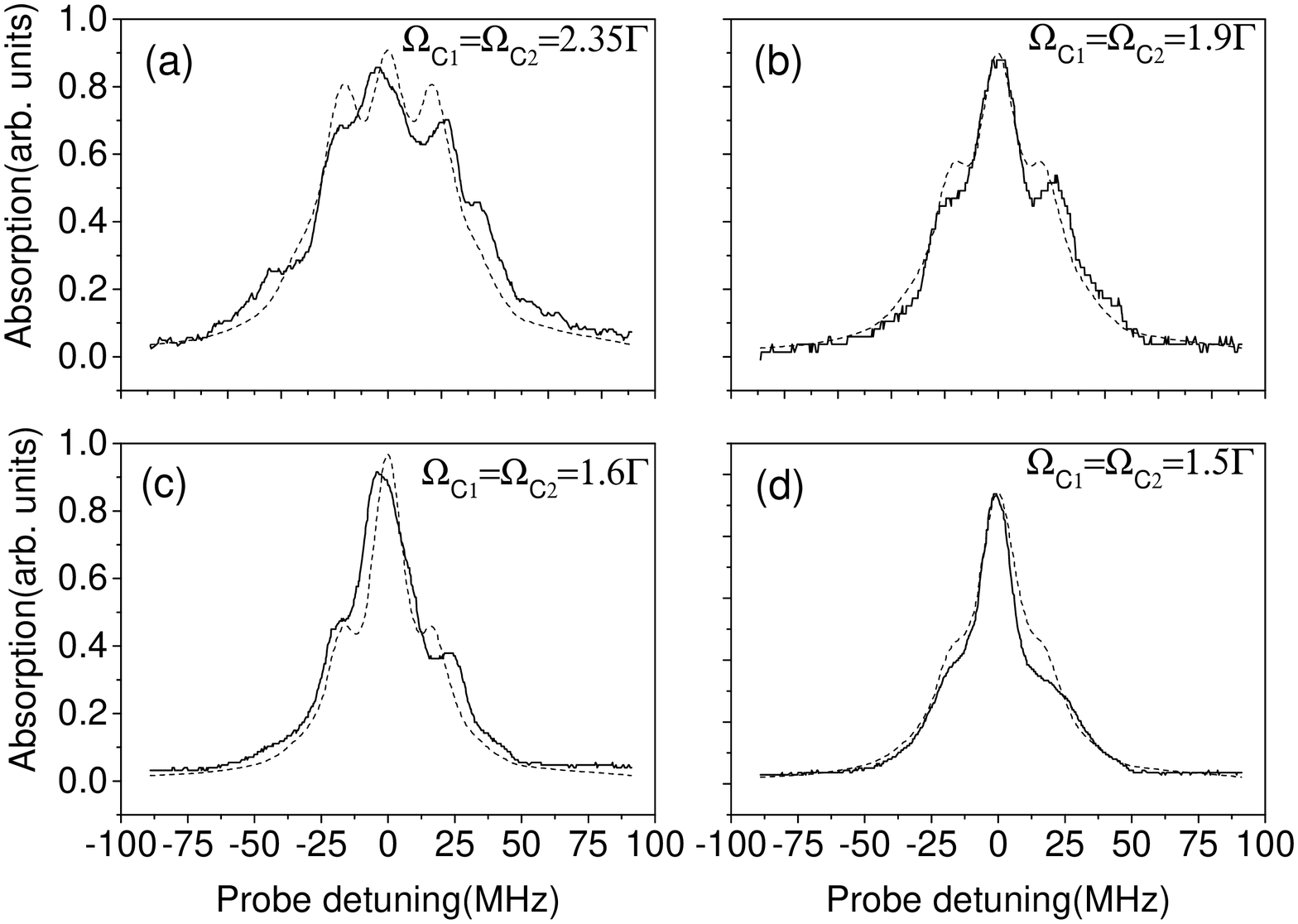,width=\columnwidth,angle=270}
\caption{Measured probe absorption versus the probe detuning in
the bichromatic coupled three-level $^{87}$Rb atoms. The solid
line is the experimental data and the dashed line is the
calculated fit. The frequency difference $2\delta=40$ MHz and the
Rabi frequency of probe field $\Omega_{p}=0.1\Gamma$. The coupling
Rabi frequencies are (a) $\Omega_{c1}=\Omega_{c2}=2.35\Gamma$
($\Gamma=6$ MHz); (b) $\Omega_{c1}=\Omega_{c2}=1.9\Gamma$, (c)
$\Omega _{c1}=\Omega_{c2}=1.6\Gamma$; and (d)
$\Omega_{c1}=\Omega_{c2}=1.5\Gamma$.}
\end{figure}

We also measured the probe absorption spectrum when the average
frequency of the bichromatic field is not resonant with the atomic
transition ($\Delta _{c2}\neq\delta$). As expected, the probe
absorption profile becomes asymmetrical. Fig. 6 shows the
experimental data (solid line) and the calculated results (dashed
line) obtained with several different values of $\Delta_{c2}$. The
coupling Rabi frequencies are $\Omega_{c1}=\Omega_{c2}=13$ MHz.
Fig. 6(a), 6(b), and 6(c) correspond to $\Delta_{c2}=20$ =
MHz$,\Delta _{c2}=34$ MHz, and $\Delta_{c2}=8$ MHz, respectively.
These measurements agree with the theoretical calculations.

\begin{figure}[tb]
\psfig{figure=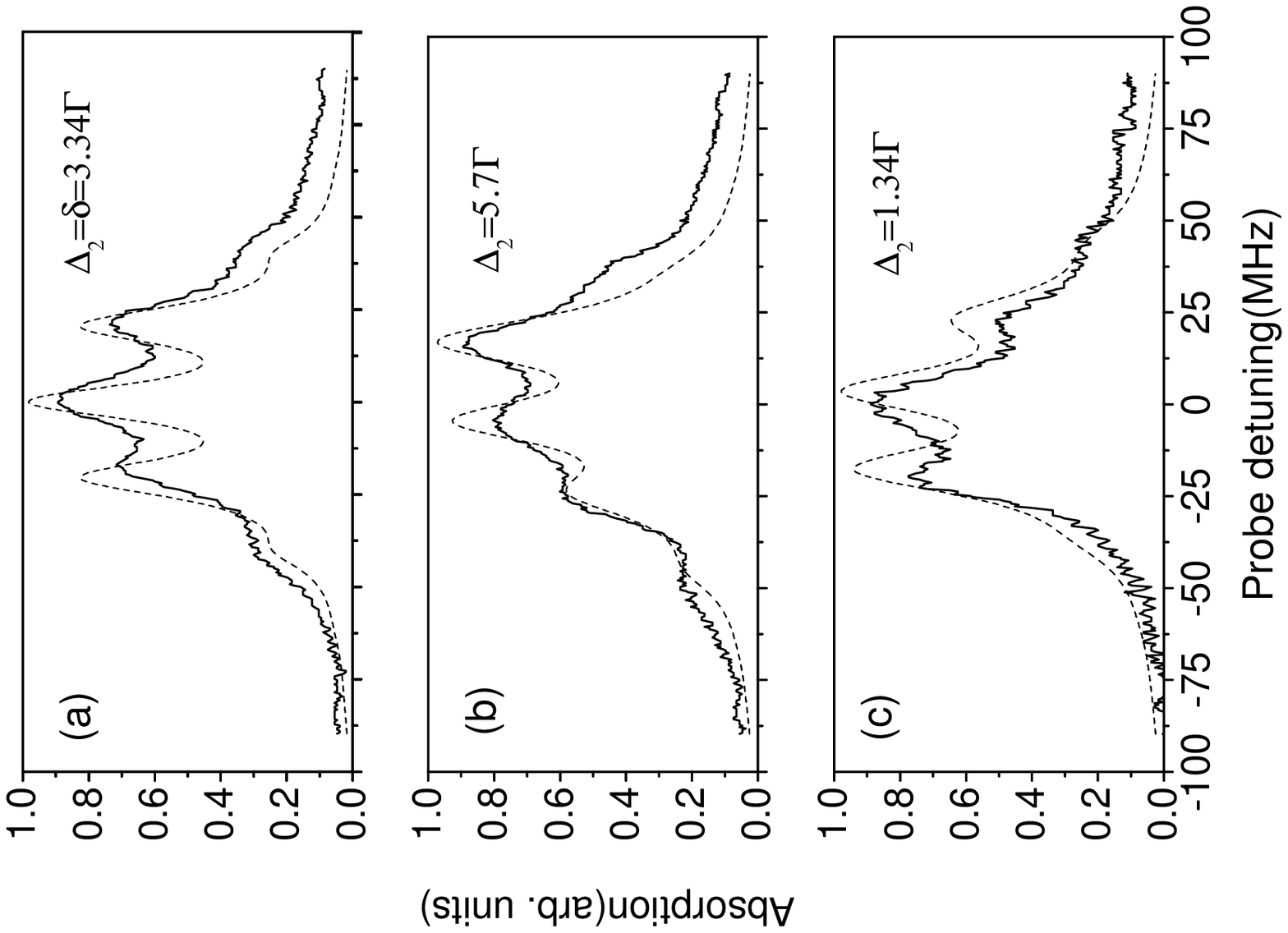,width=1.4\columnwidth,angle=270}
\caption{Measured probe absorption versus the probe detuning in
the bichromatic coupled three-level $^{87}$Rb atoms. The solid
line is the experimental data and the dash line is the calculated
fit. The Rabi frequencies of the coupling field are
$\Omega_{c1}=\Omega_{c2}=2.2\Gamma$ and the probe Rabi frequency
is $0.1\Gamma$. The frequency difference of the bichromatic field
is $2\delta=6.7\Gamma$. (a)
$\Delta_{c2}=\omega_{c}-\omega_{c1}=3.34\Gamma $(20 MHz)
($\omega_{c}$ is the frequency of the $^{87}$Rb $D_{2}$ transition
5S$_{1/2}$, $F=1\rightarrow5P_{3/2}$, $5P_{3/2},F%
\acute{}%
=2$); (b) $\Delta_{c2}=\omega_{c}-\omega_{c1}=5.7\Gamma$; and (c)
$\Delta _{c2}=1.34\Gamma$.}
\end{figure}

In summary, we have shown that a three-level $\Lambda$ system
coupled by a bichromatic field exhibits multiple absorption peaks,
which lead to the normal dispersion curves near the multiple
transparent windows. This phenomenon may be used to simultaneously
slow down photons at equally spaced frequencies, which may have
applications in quantum information processing and quantum
nonlinear optics \cite{21,22}. We have observed such bichromatic
EIT in cold $^{87}$Rb atoms. The multi-frequency transparency of
the probe light occurs under an equal-amplitude bichromatic
coupling field, which expands the frequency range of EIT and may
improve the controllability of EIT. The number of absorption peaks
depends on the intensity and the frequency separation of the
bichromatic field. The separation between two adjacent peaks
depends on the frequency difference between bichromatic coupling
lasers and is independent of the coupling intensity.

We acknowledge the financial support of National Natural Science
Foundation of China under Grant Nos. 10104018, 10074072. YZ
acknowledges support from the National Science Foundation (Grant
No. 0140032).

$\ast $Email address: wangjin@wipm.ac.cn

\end{document}